\documentclass{article}
\usepackage[latin9]{inputenc}
\usepackage{color}
\usepackage{amsmath}
\usepackage{graphicx}

\makeatletter

\providecommand{\tabularnewline}{\\}

\usepackage{spconf}

\usepackage[bookmarks=false,pdftex,pdfpagemode=None,pdfstartview=FitH,pdfview=FitH,colorlinks=true,pdftitle={Neural Speech Synthesis on a Shoestring: Improving the Efficiency of LPCNet},pdfauthor={Jean-Marc Valin, Umut Isik, Paris Smaragdis, Arvindh Krishnaswamy}]{hyperref}
\usepackage{balance}

\makeatother

\begin{document}
\name{Jean-Marc Valin, Umut Isik, Paris Smaragdis, Arvindh Krishnaswamy} \address{Amazon Web Services\\ Palo Alto, CA, USA\\ \small{\texttt{\{jmvalin, umutisik, parsmara, arvindhk\}@amazon.com}}}
\ninept
\title{Neural Speech Synthesis on a Shoestring: Improving the Efficiency
of LPCNet}

\maketitle
\maketitle 
\begin{abstract}
Neural speech synthesis models can synthesize high quality speech
but typically require a high computational complexity to do so. In
previous work, we introduced LPCNet, which uses linear prediction
to significantly reduce the complexity of neural synthesis. In this
work, we further improve the efficiency of LPCNet -- targeting both
algorithmic and computational improvements -- to make it usable on
a wide variety of devices. We demonstrate an improvement in synthesis
quality while operating 2.5x faster. The resulting open-source\footnote{Source code available at \url{https://github.com/xiph/LPCNet/} in
the \texttt{lpcnet\_efficiency} branch.} LPCNet algorithm can perform real-time neural synthesis on most existing
phones and is even usable in some embedded devices.
\end{abstract}

\section{Introduction}

\label{sec:intro}

Recent advances in neural vocoders, including WaveNet~\cite{van2016wavenet},
WaveRNN~\cite{kalchbrenner2018efficient}, and SampleRNN~\cite{mehri2016samplernn}
have demonstrated significant improvements over the capabilities of
statistical~\cite{tokuda2000speech} and concatenative~\cite{hunt1996unit}
speech synthesis. This has led to improvements in text-to-speech (TTS)~\cite{shen2018natural},
low bitrate speech coding \cite{kleijn2018wavenet}, and more. 

Unfortunately, many neural vocoders -- including most GAN\nobreakdash-~\cite{donahue2019wavegan}
and flow-based~\cite{prenger2019waveglow} algorithms -- require
a GPU for real-time synthesis, limiting their use in mobile devices.
WaveRNN was one of the first algorithms to target real-time synthesis
on a CPU. LPCNet uses linear prediction to improve the efficiency
of WaveRNN, making it possible to perform real-time synthesis on many
smartphone CPUs~\cite{valin2019lpcnet}. Even with these advances,
there is still an inherent tradeoff between synthesis quality and
complexity.

In this work, we improve on LPCNet with the goal of making it even
more efficient in terms of quality/complexity tradeoff. We propose
algorithmic improvements through hierarchical probability distribution
sampling, combined with computational improvements that seek to better
adapt LPCNet to existing CPU architectures. 

We review LPCNet and analyze its efficiency bottlenecks in Section~\ref{sec:Overview}.
We propose an efficient hierarchical sampling method in Section~\ref{sec:Model-Efficiency},
which makes it possible to increase the size of the second GRU. We
then propose improvements to the computational efficiency in Section~\ref{sec:Computational-Efficiency},
and discuss training aspects in Section~\ref{sec:Training}. In Section~\ref{sec:Results},
we evaluate the proposed improvements and demonstrate real-time synthesis
for low-power embedded platforms.

\section{LPCNet Overview}

\label{sec:Overview}

LPCNet is a proposed improvement to WaveRNN that makes use of linear
prediction to ease the task of the RNN. It operates with pre-emphasis
in the $\mu$-law domain and its output probability density function
(pdf) is used to sample a white excitation signal. Using a GRU of
size $N_{A}=384$, LPCNet can achieve high-quality real-time synthesis
with a complexity of 3~GFLOPS, using 20\% of an Intel 2.4~GHz Broadwell
core~\cite{valin2019lpcnetcodec}, a significant reduction over the
original WaveRNN (around 10~GFLOPS).

\begin{figure}
\begin{centering}
\includegraphics[width=1\columnwidth]{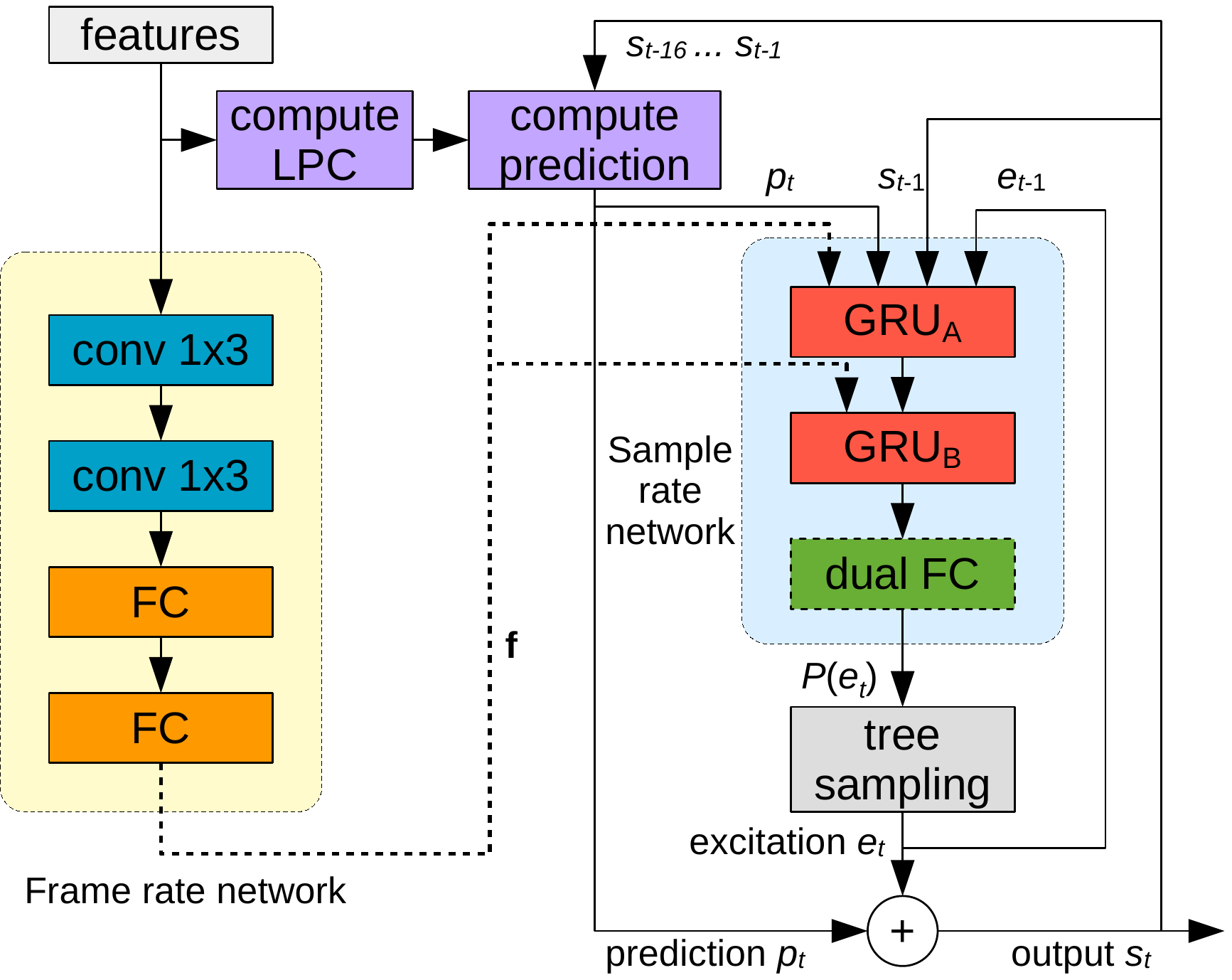}
\par\end{centering}
\caption{Overview of the improved LPCNet vocoder proposed in this work. Compared
to the original LPCNet, the softmax and direct sampling steps are
replaced by a hierarchical tree sampling. Only a subset of the dual
full-connected outputs is computed (as needed). Computational efficiency
improvements are not shown in this figure.\label{fig:Overview}}
\end{figure}

LPCNet includes both a frame rate network and a sampling rate network
(Fig.~\ref{fig:Overview}). In this paper, we focus on improving
the sampling rate network, which is responsible for more than 90\%
of the total complexity. LPCNet uses several algebraic simplifications
(Section~3.6 of~\cite{valin2019lpcnet}) to avoid the operations
related to the input matrix of the main GRU ($\mathrm{GRU_{A}})$,
making the recurrent weights responsible for most of the total complexity.
Despite that, other components contribute to the complexity, including
the second GRU (addressed in ~\cite{kanagawa2020lightweight}), as
well as the sampling process.

As we previously reported in~\cite{valin2019lpcnetcodec}, an important
bottleneck is the cache bandwidth required to load all of the sampling
rate network weights for every output sample. This is compounded by
the fact that these weights often do not fit in the L2 cache of CPUs.
A secondary bottleneck includes about 2000~activation function evaluations
per sample (for $N_{A}=384$). We propose a coherent set of improvements
that jointly alleviate most of these bottlenecks.

\section{Model Algorithmic Efficiency}

\label{sec:Model-Efficiency}

Before addressing the computational bottlenecks listed earlier, we
consider algorithmic efficiency improvements to LPCNet -- both in
terms of reducing complexity and in terms of improving quality for
a similar complexity.

\subsection{Hierarchical Probability Distribution}

\begin{figure}

\begin{centering}
\includegraphics[width=0.7\columnwidth]{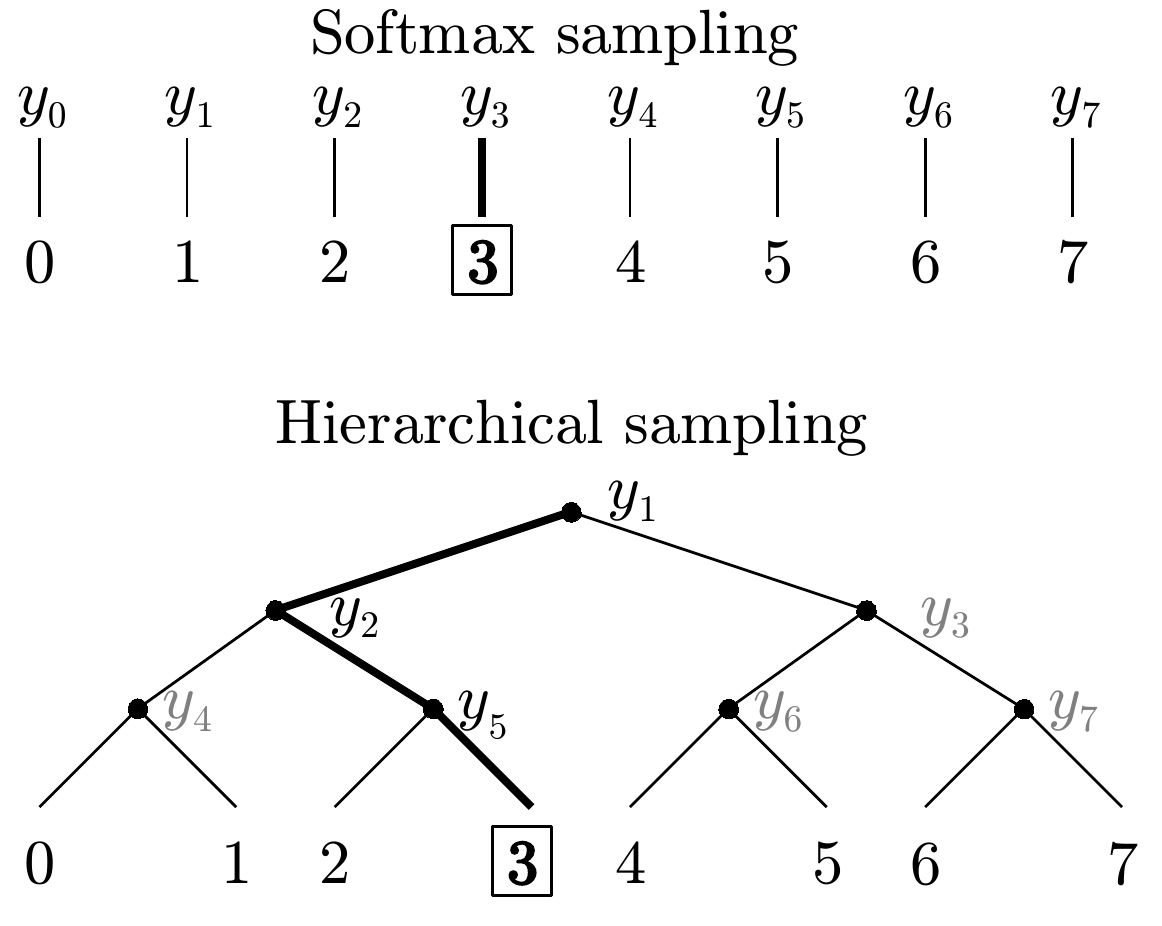}\caption{Sampling example from $Q=8$ different $\mu$-law values, with $y_{i}$
being the output logits from the dual~FC layer. With regular softmax
sampling, all 8 $y_{i}$ values need to be evaluated to compute the
partition function. With hierarchical sampling, only three $y_{i}$
values need to be evaluated (in this case, $y_{1}$, $y_{2}$, $y_{5}$)
to make the branch decisions.\label{fig:Sampling-example}}
\par\end{centering}
\end{figure}

One area where LPCNet can be improved is in sampling its output probability
distribution. In~\cite{kanagawa2020lightweight}, the authors propose
using tensor decomposition to reduce the complexity of the dual~FC
layer. In the ``bit bunching'' proposal~\cite{vipperla2020bunched},
the output resolution is extended to 11~bits by splitting the output
pdf into a 7\nobreakdash-bit softmax and an additional 4\nobreakdash-bit
softmax. In this work, we keep the resolution to 8~bits ($Q=256$),
but push the ``bunching'' idea further, splitting for every bit.
This results in the output distribution being represented as an 8\nobreakdash-level
binary tree, with each branch probability being computed as a sigmoid
output. Even though we still have 255~outputs in the last layer,
we only need to sequentially compute 8~of them when sampling, making
the sampling $O\left(\log Q\right)$ instead of $O\left(Q\right)$.
The process is illustrated in Fig.~\ref{fig:Sampling-example}.

The hierarchical probability distribution has the other benefit of
significantly reducing the number of activation function evaluations.
The original LPCNet requires 512~$\tanh\left(\right)$ evaluations
in the dual~FC layer, and 256~$\exp\left(\right)$ evaluations in
the softmax. Using hierarchical probabilities, only 16~$\tanh\left(\right)$
evaluations are required in the dual~FC layer. Moreover, the 8~$\sigma\left(\right)$
evaluations for the branch probabilities can be optimized away by
randomly sampling from a pre-computed table containing $\sigma^{-1}\left(r\right)$,
with $r\in\left]0,1\right[$.

It is often desirable to bias the sampling against low-probability
values. The original LPCNet both lowers the sampling temperature~\cite{jin2018fftnet}
on voice speech, and directly sets probabilities below a fixed threshold
to zero. With hierarchical sampling, we cannot directly manipulate
individual sample probabilities. Instead, each branching decision
is biased to render very low probability events impossible. This can
be achieved by seeding our pre-computed $\sigma^{-1}\left(r\right)$
table with $r\in\left]\xi,1-\xi\right[$. We find that $\xi=0.025$
provides natural synthesis while reducing noise and removing the occasional
glitch in the synthesis. 

\subsection{Increasing second GRU capacity}

In the original LPCNet work, $\mathrm{GRU_{B}}$ (with size $N_{B}=16$
units) acts as a bottleneck layer to avoid a high computational cost
in the following dual fully-connected layer. However, with the binary
tree described above reducing that complexity from $O\left(N_{B}Q\right)$
down to $O\left(N_{B}\log Q\right)$, it is now possible to double
$N_{B}$ without making synthesis significantly more complex. To do
that, we also make the $\mathrm{GRU_{B}}$ input matrix sparse so
that the larger $\mathrm{GRU_{B}}$ does not have too many weights
in its input matrix.

\section{Computational Efficiency}

\label{sec:Computational-Efficiency}

It is also possible to improve the efficiency of LPCNet by better
taking advantage of modern CPU architectures. Rather than making the
model mathematically smaller, these improvements make the model run
faster on existing hardware. 

\subsection{Weight Quantization}

In the original LPCNet, $\mathrm{GRU_{A}}$ uses a block-sparse floating-point
recurrent weight matrix with $16\times1$ blocks. Recently, both Intel~(x86)
and ARM introduced new ``dot product'' instructions that can be
used to compute multiple dot products of 4-dimensional 8-bit integer
vectors in parallel, accumulating the result to 32-bit integer values.
These can be used to efficiently compute products of $N\times4$ matrices
with $4\times1$ vectors. For that reason, we use block-sparse matrices
with $8\times4$ blocks, with $N=8$ being chosen as a compromise
between computational efficiency (larger $N$ is better) and sparseness
granularity (smaller $N$ is better). 

Even on CPUs where the single-instruction dot products are unavailable,
they can be emulated in a way that is still more efficient than the
use of 32-bit floating-point operations. Using 8-bit weights also
has the advantage of both dividing the required cache bandwidth by
a factor of 4 and making all of the sample rate network weights easily
fit in the L2 cache of most recent CPUs. Efficient use of the cache
is very important to avoid a load bottleneck since the sample rate
network weights are each used only once per iteration. 

To minimize the effect of quantization, we add a quantization regularization
step during the last epoch of training. We use a periodic quantization
regularization loss with local minima located at multiples of the
quantization interval $q$:
\begin{equation}
\mathcal{L}_{q}=\alpha\left(1+\epsilon-\cos\frac{2\pi w}{q}\right)^{1/4}\,,\label{eq:quant_regularization}
\end{equation}
with $\alpha=0.01$ and $\epsilon=0.001$. We use $q=1/128$, constraining
the 8-bit weights to the $\left]-1,1\right[$ range.

During the last epoch, we also gradually perform hard quantization
of the weights. Weights that are close enough to a quantization point
such that 
\begin{equation}
\left|\frac{w}{q}-\left\lfloor \frac{w}{q}\right\rceil \right|<\zeta\label{eq:hard_quant}
\end{equation}
are quantized to $q\left\lfloor \frac{w}{q}\right\rceil $, where
$\left\lfloor \cdot\right\rceil $ denotes rounding to the nearest
integer. The threshold $\zeta$ is increased linearly until $\zeta=\frac{1}{2}$,
where all weights are quantized. At that point, the sample rate network
weights are effectively frozen and only the biases and the (unquantized)
frame rate network weights are left to adjust to the quantization
process.

Weight quantization changes how the complexity is distributed among
the different layers of LPCNet, especially as the GRU size changes.
For small models, the complexity shifts away from the main GRU, making
it possible to increase the density of $\mathrm{GRU_{A}}$ without
significantly affecting the overall complexity. For large models,
the activation functions start taking an increasing fraction of the
complexity, again suggesting that we can increase the density at little
cost.

\subsection{Hyperbolic Tangent Approximation}

\begin{table}
\caption{Rational hyperbolic tangent approximation coefficients.\label{tab:Rational-approximation}}

\begin{centering}
\vspace{0.5em}
\par\end{centering}
\centering{}%
\begin{tabular}{clcl}
\hline 
$N_{0}$ & 1565.0352 & $D_{0}$ & 1565.3572\tabularnewline
$N_{1}$ & 158.3758 & $D_{1}$ & 679.1774\tabularnewline
 &  & $D_{2}$ & 19.5291\tabularnewline
\hline 
\end{tabular}
\end{table}

As computational complexity is reduced through weight quantization
and the use of sparse matrices, computing the activation functions
becomes an increasingly important fraction of the total complexity.
For that reason, we need a more efficient way to compute the sigmoid
and hyperbolic tangent functions. We do not consider methods based
on lookup tables since those are usually hard to vectorize, and although
methods based on exponential approximations are viable, we are seeking
an even faster method using a direct $\tanh\left(\cdot\right)$ approximation.
For those reasons, we consider the Padé-inspired clipped rational
function
\begin{equation}
\tilde{\tau}\left(x\right)=\mathrm{clip}\left(x\cdot\frac{N_{0}+N_{1}x^{2}+x^{4}}{D_{0}+D_{1}x^{2}+D_{2}x^{4}},\,-1,\,1\right)\,.\label{eq:tanh_approx}
\end{equation}

We optimize the $D_{k}$ and $N_{k}$ coefficients by gradient descent
to minimize the maximum error $E=\max_{x}\left|\tilde{\tau}\left(x\right)-\tanh\left(x\right)\right|$
and find that the coefficients in Table~\ref{tab:Rational-approximation}
result in $E=6\cdot10^{-5}$. Using Horner's method for polynomial
evaluation, the approximation in \eqref{eq:tanh_approx} can be implemented
using 10~arithmetic instructions on both x86 (AVX/FMA) and ARM (NEON)
architectures. Instead of the division, we use the hardware reciprocal
approximation instructions, resulting in a final accuracy of $\pm3\cdot10^{-4}$
on x86. The error is roughly uniformly distributed, except for large
input values, for which the output is exactly equal to $\pm1$.

Since $\sigma\left(x\right)=\frac{1}{2}+\frac{1}{2}\tanh\frac{x}{2}$,
the sigmoid function can be similarly approximated -- still using
10~arithmetic instructions -- by appropriately scaling the polynomial
coefficients and adding an offset:
\begin{equation}
\tilde{\sigma}\left(x\right)=\mathrm{clip}\left(\frac{1}{2}+x\cdot\frac{16N_{0}+4N_{1}x^{2}+x^{4}}{64D_{0}+16D_{1}x^{2}+4D_{2}x^{4}},\,0,\,1\right)\,.\label{eq:sigmoid_approx}
\end{equation}
Again, the approximation exactly equals 0 or 1 for large input values.
That is an important property for use in gated RNNs, as it allows
perfect retention of the state when needed, while also avoiding the
exponential growth that could occur if the gate value exceeded unity.

\section{Training}

\label{sec:Training}

The training procedure is similar to the one described in~\cite{valin2019lpcnetcodec},
with Laplace-distributed noise injected in the $\mu$-law excitation
domain. To ensure robustness against unseen recording environments,
we apply random spectral augmentation filtering using a second-order
filter, as described in Eq.~(7) of~\cite{valin2018rnnoise}. 

Like other auto-regressive models, LPCNet relies on teacher forcing~\cite{williams1989teacher}
and is subject to exposure bias~\cite{bengio2015scheduled,ranzato2016sequence}.
The phenomenon becomes worse for smaller models since the decreased
capacity limits the use of regularization techniques such as early
stopping. We find that initializing the RNN state using the state
from a random previous sequence helps the network generalize to inference.
That can be easily accomplished using Tensorflow's \emph{stateful}
RNN option, while still randomizing the training sequence ordering. 

\section{Experiments and Results}

\label{sec:Results}

We evaluate the proposed LPCNet improvements on a speaker-independent,
language-independent synthesis task where the inputs features are
computed directly from a reference speech signal. All models are trained
using 205~hours of 16-kHz speech from a combination of TTS datasets~\cite{demirsahin-etal-2020-open,kjartansson-etal-2020-open,kjartansson-etal-tts-sltu2018,guevara-rukoz-etal-2020-crowdsourcing,he-etal-2020-open,oo-etal-2020-burmese,van-niekerk-etal-2017,gutkin-et-al-yoruba2020,bakhturina2021hi}
including more than 900~speakers in 34~languages and dialects. To
make the data more consistent, we ensure that all training samples
have a negative polarity. This is done by estimating the skew of the
residual, in a way similar to~\cite{drugman2013residual}. 

For all models, we use sequences of 150~ms (15~frames of 10~ms)
and a batch size of 128~sequences. The models are trained for 20~epochs
(767k~updates) using Adam~\cite{kingma2014adam} with $\beta_{1}=0.9$
and $\beta_{2}=0.99$ and a decaying learning rate $\alpha=\frac{\alpha_{0}}{1+\delta\cdot b},$where
$\alpha_{0}=0.001,$ $\delta=5\times10^{-5}$ and $b$ is the update
number. The sparse weights are obtained using the technique described
in~\cite{kalchbrenner2018efficient}, with the schedule starting
at $b=2,000$ and ending at $b=40,000$. As in~\cite{valin2019lpcnetcodec},
for an overall weight density $d$ in a GRU, we use $2d$ for the
state weights and $d/2$ for both the update and reset gates. 

We compare the proposed improvements to a baseline LPCNet~\cite{valin2019lpcnet}
for different GRU sizes. The GRU sizes, $N_{A}$ and $N_{B}$, and
the weight density, $d_{A}$ and $d_{B}$, are listed in Table~\ref{tab:Model-sizes},
with the models named for \emph{baseline} (B) or \emph{proposed} (P),
followed by the $\mathrm{GRU_{A}}$ size. For example, P384 is the
proposed model for $N_{A}=384$. 

\begin{table}
\begin{centering}
\caption{Definition of the models being evaluated, including the number of
weights \emph{used} (one multiply-add per weight) for each iteration
of the sample rate network. \label{tab:Model-sizes}}
\vspace{0.5em}
\par\end{centering}
\centering{}%
\begin{tabular}{ccccccc}
\hline 
Model & $N_{A}$ & $d_{A}$ & $N_{B}$ & $d_{B}$ & Quantized & Weights\tabularnewline
\hline 
B192 & 192 & 0.1 & 16 & dense & no & 30k\tabularnewline
B384 & 384 & 0.1 & 16 & dense & no & 73k\tabularnewline
B640 & 640 & 0.1 & 16 & dense & no & 165k\tabularnewline
\hline 
\textbf{P192} & 192 & 0.25 & 32 & 0.5 & yes & 40k\tabularnewline
\textbf{P384} & 384 & 0.1 & 32 & 0.5 & yes & 66k\tabularnewline
\textbf{P640} & 640 & 0.15 & 32 & 0.5 & yes & 219k\tabularnewline
\hline 
\end{tabular}
\end{table}

\subsection{Complexity}

\begin{table}[!t]
\caption{Complexity of the baseline and proposed models on x86, Neoverse N1,
Cortex-A72, and Cortex-A53. The complexity is expressed as a percentage
of the core compute time required for real-time synthesis (inverse
of the real-time factor). A value below 100\% means that the model
can run in real-time on a particular core. Values in red indicate
that real-time synthesis is not possible on the corresponding CPU.
The speedup factor is computed as the geometric mean of the ratio
over the three model sizes.\label{tab:Complexity-of-models}}

\begin{centering}
\vspace{0.5em}
\par\end{centering}
\begin{centering}
\begin{tabular}{ccccc}
\hline 
Model & x86 (\%) & N1 (\%) & A72 (\%) & A53 (\%)\tabularnewline
\hline 
B192 & 6.1 & 15.7 & 59 & \textcolor{red}{159}\tabularnewline
B384 & 13.2 & 28.5 & \textcolor{red}{108} & \textcolor{red}{277}\tabularnewline
B640 & 29.5 & 55 & \textcolor{red}{224} & \textcolor{red}{740}\tabularnewline
\hline 
\textbf{P192} & 2.8 & 7.1 & 38 & 92\tabularnewline
\textbf{P384} & 4.5 & 11.8 & 65 & \textcolor{red}{154}\tabularnewline
\textbf{P640} & 11.6 & 26.0 & \textcolor{red}{150} & \textcolor{red}{350}\tabularnewline
\hline 
Speedup & 2.5x & 2.2x & 1.6x & 1.9x\tabularnewline
\hline 
\end{tabular}
\par\end{centering}
\begin{centering}
\caption{Results from the MOS quality evaluation on both test sets, as well
as the average of the two sets (overall). The confidence interval
(CI) on the overall quality is 0.03 for all algorithms, except for
the reference that has a CI of 0.02. All the differences in overall
results are statistically significant, with $p<.01$.\label{tab:MOS-results}}
\vspace{0.5em}
\par\end{centering}
\centering{}%
\begin{tabular}{cccc}
\hline 
Model & PTDB-TUG & NTT & \textbf{Overall}\tabularnewline
\hline 
Reference & 4.19 & 4.27 & 4.23\tabularnewline
Speex 4k & 2.61 & 2.75 & 2.68\tabularnewline
\hline 
B192 & 3.63 & 3.65 & 3.64\tabularnewline
B384 & 3.92 & 4.00 & 3.96\tabularnewline
B640 & 3.99 & 4.11 & 4.05\tabularnewline
\hline 
\textbf{P192} & 3.76 & 3.87 & 3.81\tabularnewline
\textbf{P384} & 3.93 & 4.07 & 4.00\tabularnewline
\textbf{P640} & 4.03 & 4.17 & 4.10\tabularnewline
\hline 
\end{tabular}
\end{table}

We evaluate the complexity of the proposed improvements on four different
cores: an Intel i7-10810U mobile x86 core, a 2.5~GHz ARM Neoverse
N1 core with similar single-core performance as recent smartphones,
a 1.5~GHz ARM Cortex-A72 core similar to older smartphones, and a
1.4~GHz ARM Cortex-A53 core as found in some embedded systems.

The baseline and proposed models are implemented in C and share most
of the code. Both use the same amount of hand-written AVX2 and NEON
intrinsics to implement the DNN models. The measured complexity of
the different models on each of the four cores is shown in Table~\ref{tab:Complexity-of-models}.
The measurements show that the computational requirement on x86 is
reduced by a factor of 2.5x. In addition, the improved medium-sized
model can now operate in real time on older phones ($\sim$2016),
whereas the smaller model can operate on some existing low-power embedded
systems.

\subsection{Quality}

We evaluate the models on the PTDB-TUG speech corpus~\cite{PirkerWPP11}
and the NTT Multi-Lingual Speech Database for Telephonometry. From
PTDB-TUG, we use all English speakers (10~male, 10~female) and randomly
pick 200~concatenated pairs of sentences. For the NTT database, we
select the American English and British English speakers (8~male,
8~female), which account for a total of 192~samples (12~samples
per speaker). The training material did not include any data from
the datasets used in testing. In addition to the 6~models described
in Table~\ref{tab:Model-sizes}, we also evaluate the reference speech
as an upper bound on quality, and we include the Speex 4~kb/s wideband
vocoder~\cite{valin2007speex} as an anchor.

\begin{figure}[t]
\begin{centering}
\includegraphics[width=1\columnwidth]{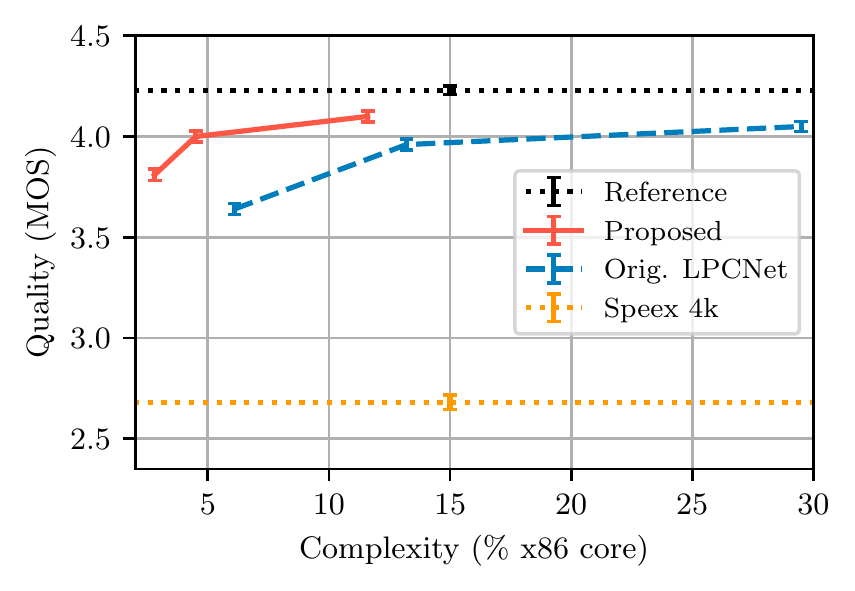}
\par\end{centering}
\caption{Synthesis quality as a function of the complexity. We plot the x86
complexity from Table~\ref{tab:Complexity-of-models} with the overall
quality from Table~\ref{tab:MOS-results}. The two curves approximately
match when we scale the computational complexity of one of them by
3.5. This means that to obtain equal quality, the proposed method
reduces the computational complexity by a factor of 3.5x. \label{fig:MOS-vs-complexity}}
\end{figure}

The mean opinion score (MOS)~\cite{P.800} results were obtained
using the crowdsourcing methodology described in P.808~\cite{P.808}.
Each file was evaluated by 20~randomly-selected listeners. The results
in Table~\ref{tab:MOS-results} show that for the same $\mathrm{GRU_{A}}$
size, the proposed models all perform significantly better than the
baseline LPCNet models. The results are also consistent across the
two datasets evaluated. This suggests that the small degradation in
quality caused by weight quantization is more than offset by the increase
in $N_{B}$ and (for P192 and P640) the density increase. Figure~\ref{fig:MOS-vs-complexity}
illustrates how the proposed models affect the LPCNet quality-complexity
tradeoff.

\section{Conclusion}

\label{sec:conclusion}

We have proposed algorithmic and computational improvements to LPCNet.
We demonstrate speed improvements of 2.5x or more, while providing
better quality than the original LPCNet, equivalent to an efficiency
improvement of 3.5x at equal quality. The proposed improvements make
high-quality neural synthesis viable on low-power devices. Considering
that the proposed changes are independent of previously proposed enhancements,
such as multi-sample or multi-band sampling, we believe further improvements
to the LPCNet efficiency are possible. 

\balance

\bibliographystyle{IEEEbib}
\bibliography{lpcnet,corpora}

\end{document}